\documentstyle[11pt,psfig]{l-aa}

\def\farcm{\hbox{$.\!\!^{\prime}$}}
\def\farcs{\hbox{$.\!\!^{\prime\prime}$}}  % Fractions of arcseconds
\def\sss{\hbox{$.\!\!^{s}$}}

\def\asec{\ifmmode ^{\prime\prime}\else$^{\prime\prime}$\fi}
\begin{document}
\thesaurus{06.			% A&A Section 6: Formation, structure and evolution of stars
		(08.14.1;	% Stars: neutron
		08.16.7;        % Stars: pulsars: individual
		03.20.4;	% Techiques: photometric
		03.20.1)	%Techniques: image processing
		}
\title{
BVRI observations of PSR~B0656+14 with the 6-meter telescope.
}
\author{V. G. Kurt \inst{1}, V. V. Sokolov \inst{2}, S. V. Zharikov \inst{2}, G. G. Pavlov \inst{3} and  B.V. Komberg \inst{1}}
\institute{$^1$Astro Space Center of the Russian Academy of Sciences, 117810, Moscow,
Russia;  vkurt@dpc.asc.rssi.ru  \\
$^2$Special Astrophysical Observatory of RAS,
Karachai-Cherkessia, Nizhnij Arkhyz, 357147 Russia; sokolov@sao.ru \\
$^3$The Pennsylvania State University,
Dept.~of Astronomy \& Astrophysics, 535 Davey Lab, 
University Park, PA 16802; pavlov@astro.psu.edu}
\date{\today}
\maketitle
\markboth{V. G. Kurt et al.: BVRI 
observations of PSR~B0656+14}{}

\begin{abstract}
We observed the middle-aged radio pulsar B0656+14
with a CCD detector at the 6-m telescope.
Broadband {\it BVRI} images
show the following magnitudes of the pulsar
counterpart:
$B = 24.85 (+0.19, -0.16),\
V = 24.90 (+0.16, -0.14),\
R = 24.52 (+0.12, -0.11),\
I = 23.81 (+0.27, -0.21)$.
We fitted the UV-optical (space + ground-based) data with a
two-component model which combines
a power law (non-thermal component)
with a thermal spectrum emitted by the neutron star surface.
The power law component, with the
 energy power-law index $\alpha=1.5 (+1.1,-1.2)$,
dominates in the observed range. Constraints on the thermal component
correspond to the Rayleigh-Jeans parameter $G\equiv T_6(R_{10}/d_{500})^2
=4.1 (+2.1, -4.1)$, where $T=10^6T_6$~K is the brightness temperature,
$R_\infty = 10 R_{10}$~km is the neutron star radius as seen by
a distant observer, and $d=500 d_{500}$~pc is the distance.
The shape of the optical-UV spectrum of PSR~B0656+14 differs considerably from
those observed from other pulsars:
the middle-aged Geminga and young Crab, Vela.

% ground-based observations, CCD photometry}
\end{abstract}
% *************************************************************************
%                            Author's own macros
% *************************************************************************
\newcommand{\gapr}{\raisebox{-.6ex}{\mbox{
$\stackrel{>}{\mbox{\scriptsize$\sim$}}\:$}}}
\newcommand{\lapr}{\raisebox{-.6ex}{\mbox{
$\stackrel{<}{\mbox{\scriptsize$\sim$}}\:$}}}

\section{ Introduction}

Optical radiation has been detected
from only $\sim 1\%$
of $\sim 750$ radio pulsars
(see Bignami \& Caraveo 1996
 for a recent review).
Optical counterparts
were reported for pulsars of quite different ages,
from the very young PSR B0531+21
(Crab pulsar --- see \cite{12})
to old PSR B1929+10 and B0950+08 (Pavlov, Stringfellow \&
Cordova 1996a). Of special
interest among these objects are the middle-aged
($\tau\sim 10^5-10^6$~yr), soft X-ray pulsars
PSR~B0656+14, J0633+1746 (Geminga) and B1055--52
studied extensively with the X-ray observatories
$ROSAT$ and $ASCA$. Their thermal-like
X-ray radiation is believed to be emitted by surface layers of
the
neutron stars (e.~g., \"Ogelman 1995), so that
studying this radiation
enables one to estimate surface temperatures of
neutron stars (NSs) of different ages, to compare them with
the current models for the NS cooling, and to
constrain properties of the superdense matter in the
NS interiors (e.~g., Pavlov et al. 1995).
Since the surface temperature, as well as other parameters
of radiating layers, can be measured much more accurately
if the surface radiation is also observed in the optical--UV
range
(Pavlov et al. 1996b),
investigation of the optical--UV radiation of these objects
is potentially very important.

In this paper we report multi-color CCD observations of the middle-aged
PSR~B0656+14 ($\tau = P/2\dot{P}=1.1\times 10^5$~yr)
 with the 6-meter telescope (BTA) at the Special Astrophysical
Observatory of the Russian Academy of Sciences (SAO RAS).
The optical counterpart of this pulsar
 was first detected with the ESO NTT and 3.6-meter
telescopes in the $V$ band ($V\sim 25$)
by Caraveo et al.~(1994a) at a 3$\sigma$ level. Pavlov et al. (1996a)
observed the pulsar with the Faint Object Camera (FOC)
onboard the Hubble Space Telescope ($HST$) in a longpass filter F130LP
($\lambda\lambda = 2310-4530$~\AA).
They confirmed the optical detection at a very high confidence
level ($S/N=52$) and concluded that the observed radiation is
mainly of a nonthermal (most likely, magnetospheric) origin, although a
contribution of a thermal (Rayleigh--Jeans) component emitted
by the NS surface cannot be excluded.
In order to separate
the thermal component, and to understand the nature of the
nonthermal radiation, the flux of the object should be
measured in several
spectral bands covering as broad spectral range as possible.
Mignani, Caraveo \& Bignami (1997) observed PSR~B0656+14 with the
$HST$ Planetary Camera (PC) in the F555W filter and determined
its $V$ magnitude more accurately: $V=25.1\pm 0.1$.
Obviously, observations with other filters are needed
to find the shape of the spectral flux of the optical
counterpart.
Thus, we included such observations in our
general program of deep broad-band CCD photometry
of nearby pulsars of the northern sky with the 6-meter telescope.
\begin{figure*}
   \centerline{
    \vbox{\psfig{figure=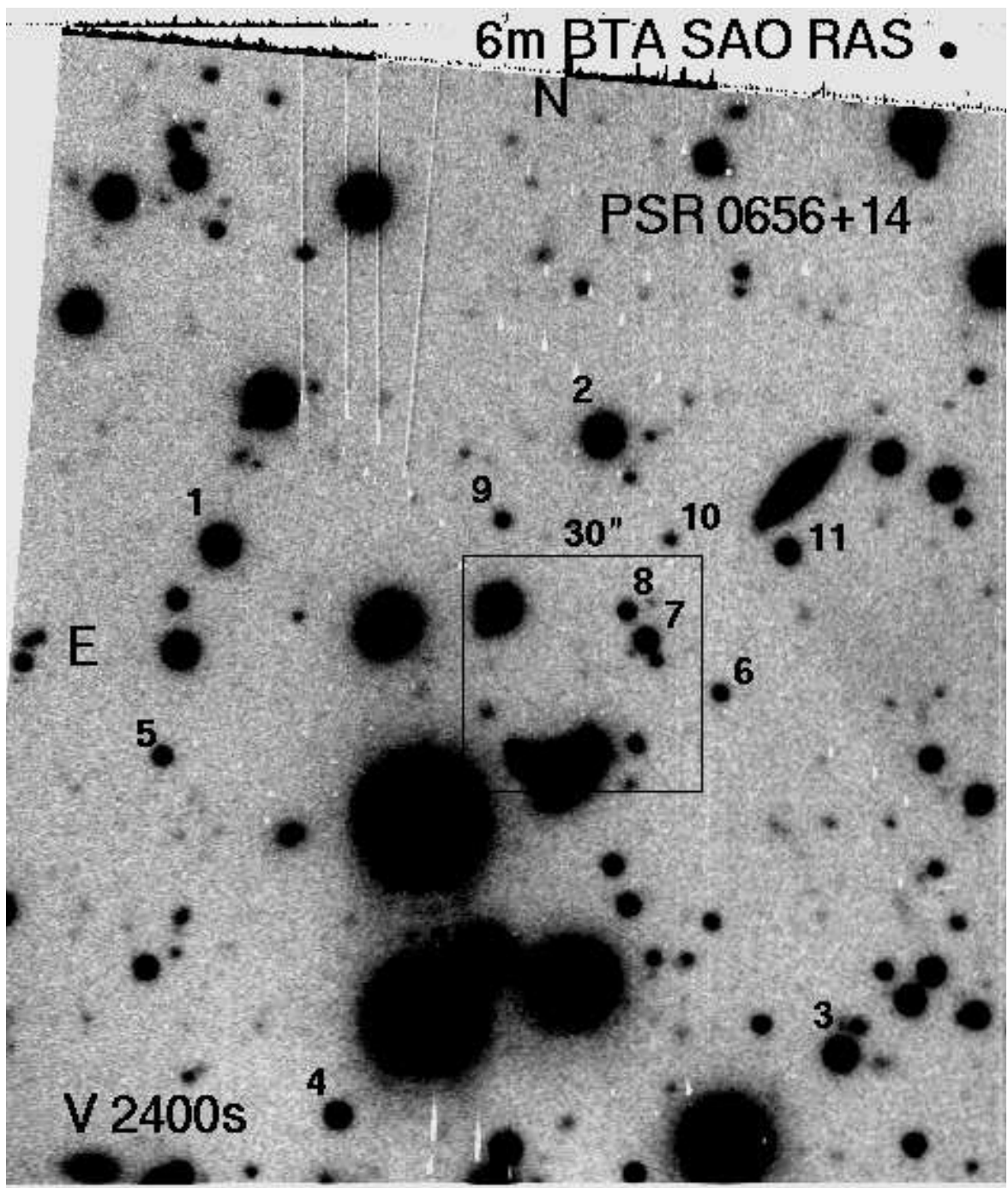,width=15.5cm,%}}\par
	bbllx=90pt,bblly=320pt,bburx=450pt,bbury=750pt,clip=}}
}
\caption{ Image of the PSR~B0656+14 field in the $V$ filter
with accumulation time of
2400 sec.
The center of the $30^{\prime\prime}\times30^{\prime\prime}$ fragment
coincides with the pulsar position.
The numbers denote the stars whose magnitudes are given in Table 3.
}
\label{field}
 \end{figure*}
%%%%%%%%%%%%%%%%%%%%%
\begin{table}
\caption[ ]{Observations of PSR~B0656+14 on November 11/12, 1996.}
\begin{center}
\begin{tabular}{cccccl}
\hline
 time & exp.  &  exposure   &  seeing     & $Z$       &  backgr.     \\
 UT   & ID    &             &             &           & (mag/        \\
      &         & (sec)      &(arcsec)    &  (deg)    &arcsec$^2$) \\ \hline
22:37 &   B  1  &  600        &   1.64      &  40.63     &  22.41        \\
22:49 &   V  1  &  600        &   1.56      &  39.00     &  21.60        \\
23:01 &   R  1  &  600        &   1.42      &  37.43     &  20.70        \\
23:11 &   I  1  &  300        &   1.31      &  35.95     &  19.39        \\
23:18 &   I  2  &  300        &   1.36      &  35.17     &  19.34        \\
23:25 &   B  2  &  600        &   1.69      &  34.42     &  22.39        \\
23:36 &   V  2  &  600        &   1.42      &  33.21     &  21.61        \\
23:47 &   R  2  &  600        &   1.31      &  32.13     &  20.80        \\
23:59 &   I  3  &  300        &   1.42      &  31.22     &  19.52        \\
00:06 &   I  4  &  300        &   1.33      &  30.76     &  19.47        \\
00:12 &   I  5  &  300        &   1.26      &  30.39     &  19.45        \\
00:19 &   B  3  &  600        &   1.67      &  30.06     &  22.34        \\
00:30 &   V  3  &  600        &   1.31      &  29.65     &  21.52        \\
00:42 &   R  3  &  600        &   1.45      &  29.43     &  20.64        \\
01:13 &   R*4  &  600        &   1.45      &  29.88     &  20.64        \\
01:25 &   I  6  &  300        &   1.37      &  30.42     &  19.29        \\
01:31 &   I  7  &  300        &   1.40      &  30.50     &  19.32        \\
01:41 &   I  8  &  300        &   1.30      &  31.57     &  19.41        \\
01:48 &   B  4  &  600        &   1.75      &  32.10     &  22.27        \\
01:59 &   V  4  &  600        &   1.61      &  33.18     &  21.41        \\
02:11 &   R  5  &  600        &   1.51      &  34.39     &  20.57        \\
02:22 &   R  6  &  600        &   1.53      &  35.74     &  20.53        \\
02:34 &   I  9  &  300        &   1.51      &  37.20     &  19.22        \\
02:41 &   I  10 &  300        &   1.53      &  38.05     &  19.05        \\
02:47 &   I  11 &  300        &   1.51      &  38.95     &  18.75        \\
\end{tabular}
\end{center}
\label{Zhyr}
\end{table}

In our first BTA observations
of PSR~B0656+14 (January 1996) in the {\it B, V} and {\it R}
filters (Kurt et al. 1997a)
we obtained
estimates of its brightness in the $V$ and, for the first time,
{\it B} and $R$ bands and confirmed the nonthermal nature
of its optical spectrum. However, because of poor weather conditions
during those observations, the fluxes were determined with substantial
uncertainties. Subsequent observations of the same object (November 1996)
were more successful (Kurt et al.~1997b).
We describe these observations and their
reduction in Section 2 and discuss the properties and
origin of the optical spectrum
of PSR~B0656+14 in Section 3.

\section {Observations and data analysis}

\subsection{Observations}

Photometric observations of PSR~B0656+14
were carried out with the
6-meter
telescope at SAO RAS on 11/12 November, 1996 with a CCD photometer
installed at the Primary Focus.
We used the ``Electron ISD017A'' CCD; its format of  $1040\times1160$ pixels
corresponds to the field of view
$2\farcm38\times2\farcm66$.
The CCD photometer
was employed in the $2\times2$ binning mode,
so that each of the
$520\times580$ zoomed pixels (referred to as `pixels' hereafter)
has the angular size of
$0\farcs274 \times 0\farcs274$.
We used the gain of $2.3 e^-$ per DN (Data Number)  pixel.
The shutter timing accuracy is better than 0.1 sec.

The observations were carried out with filters
close to the {\it B, V, R} and $I$ filters of the
Cousins system. Table 1 gives basic parameters of observational
conditions: starting time of each exposure, filter and exposure
number,
duration of exposure, seeing, zenith distance $Z$,
and sky background in stellar magnitudes
per arcsec$^2$. \\

\begin{table}
\begin{center}
\caption[ ]{Coordinates of objects in the PSR~B0656+14 field.}
\begin{tabular}{ccccc}
\hline
No.   & $\alpha_{1950}$ & $\delta_{1950}$ & relative   & distance   \\
     &               &                  & coordinates   &  from PSR   \\
     &                 &                & (arcsec) &     (arcsec)      \\
\hline
1    & 06:57:00.97     & 14:18:51.5     &            & \\
2    & 06:56:57.62     & 14:19:05.1     &            & \\
3    & 06:56:55.63     & 14:17:43.6     &            & \\
4    & 06:57:00.02     & 14:17:36.6     &            &  \\
5    & 06:57:01.51     & 14:18:24.1     &            &  \\
6    & 06:56:56.63     & 14:18:31.3     &            &  \\
o1,PSR & 06:56:57.84   & 14:18:34.2     &  206.5;\ 203.0& 0.0   \\
o2   & 06:56:57.75     & 14:18:36.2     &  207.5;\ 204.3& 1.7    \\
o3   & 06:56:57.87     & 14:18:37.4     &  206.1;\ 207.1& 4.1      \\
o4   & 06:56:57.47     & 14:18:34.7     &  212.0;\ 204.0& 5.1      \\
o5   & 06:56:57.58     & 14:18:34.7     &  209.5;\ 204.0& 3.1       \\  \hline
\end{tabular}
\label{coord}
\end{center}
\end{table}

\subsection{Data reduction}

Standard data reduction includes
subtraction of the ``bias'',
--- an additional component of the CCD  signal,
flat-fielding (correction for non-uniform sensibility
of the detector
% receiver
elements), and removing of space particle traces.
The data were processed with the use of the MIDAS software.
Processing involved all the obtained images except
for R*4
 where the space particle
trace was in the immediate vicinity of the pulsar position. Figure 1
shows the field around the pulsar
(sum of 4 exposures in the $V$ filter).

\subsection{Astrometrical referencing}

According to
Thompson and Cordova (1994), the VLA position of the pulsar
is $\alpha_{1950}$ = $06^h\ 56^m\ 57\sss 942$;
$\delta_{1950}=14^\circ 18^{\prime} 33\farcs 80$ (epoch 1992.98);
the formal uncertainties of the radio position ($0\farcs 09$ and
$0\farcs 03$ for $\alpha$ and $\delta$, respectively) do
not include possible systematic errors which may be as large as
$\sim 0\farcs 2$
(cf. Pavlov et al.~1996a). With allowance
for the proper motion of the pulsar, $\mu_\alpha=+0.73\pm 20$
and $\mu_\delta=-26\pm 13$ milliarcsec yr$^{-1}$ (Pavlov et
al.~1996a), the expected position of the pulsar at the epoch
of our observations (1996.86) is $\alpha_{1950}$ = $06^h\ 56^m\ 57\sss
96 (\pm 0\sss 02)$, $\delta_{1950}=14^\circ 18^{\prime} 33\farcs 7
(\pm 0\farcs 2)$.

The position of the candidate optical counterpart
relative to nearest bright stars found with the use of
the $HST$ Guige Star Catalog (GSC) has been presented
by Caraveo et. al. (1994b). The astrometric referencing of our data
is based on coordinates taken for the field stars around the pulsar
from the Digitized Sky Servey (nominal accuracy $\sim 1^{\prime\prime}$).

 Table 2 gives
the coordinates of the pulsar optical candidate, 6 bright
stars (denoted by the numbers 1 through 6 in Fig.~1), and 4 objects
within $\sim 5\asec$ around the pulsar.
For these objects,
the coordinates corresponding to the frames of
Figs.~3, 4 and 6 are also presented.

\begin{table*}
\begin{center}
\caption{Stellar magnitudes of 11
stars around the
pulsar.}
\begin{tabular}{cccccc}
\hline
   &  $B$                       &  $V$        & $R$         & $I$                       \\
    & ($\lambda_{\rm eff}=4448$~\AA,  & ($\lambda_{\rm eff}=5505$~\AA, &
      ($\lambda_{\rm eff}=6588$~\AA, & ($\lambda_{\rm eff}=8060$~\AA, \\
    &    FWHM=1008~\AA)  &  FWHM=827~\AA) &  FWHM=1568~\AA) &  FWHM=1542~\AA)           \\ \hline
  1 &    $19.35\pm0.01$         &   $18.31\pm0.01$             &  $17.72\pm0.01$             &  $17.17\pm0.01$                  \\
  2 &    $18.71\pm0.01$         &   $18.02\pm0.01$             &  $17.66\pm0.01$             &  $17.32\pm0.01$                  \\
  3 &    $19.90\pm0.01$         &   $18.69\pm0.01$             &  $17.96\pm0.01$             &  $17.36\pm0.01$                   \\
  4 &    $20.94\pm0.02$         &   $19.84\pm0.01$             &  $19.19\pm0.01$             &  $18.64\pm0.01$                   \\
  5 &    $22.61\pm0.02$         &   $21.31\pm0.01$             &  $20.54\pm0.01$             &  $19.93\pm0.01$                   \\
  6 &    $22.70\pm0.03$         &   $21.96\pm0.02$             &  $21.52\pm0.01$             &  $21.12\pm0.02$                   \\
  7 &    $21.99\pm0.03$        &   $20.31\pm0.02$             &  $19.13\pm0.01$             &  $17.74\pm0.02$                   \\
  8 &    $23.08\pm0.03$         &   $21.52\pm0.02$             &  $20.50\pm0.01$             &  $19.53\pm0.02$                   \\
  9 &    $23.58\pm0.03$         &   $22.06\pm0.02$             &  $21.24\pm0.01$             &  $20.44\pm0.02$                   \\
  10 &   $24.52\pm0.03$         &   $22.88\pm0.02$             &  $21.92\pm0.01$             &  $21.08\pm0.02$                   \\
  11 &   $21.38\pm0.03$         &   $20.40\pm0.02$             &  $19.89\pm0.01$             &  $19.39\pm0.02$                   \\ \hline
\end{tabular}
\end{center}
\label{Filt}
\end{table*}

\begin{figure*}[t]
   \centering{
    \vbox{\psfig{figure=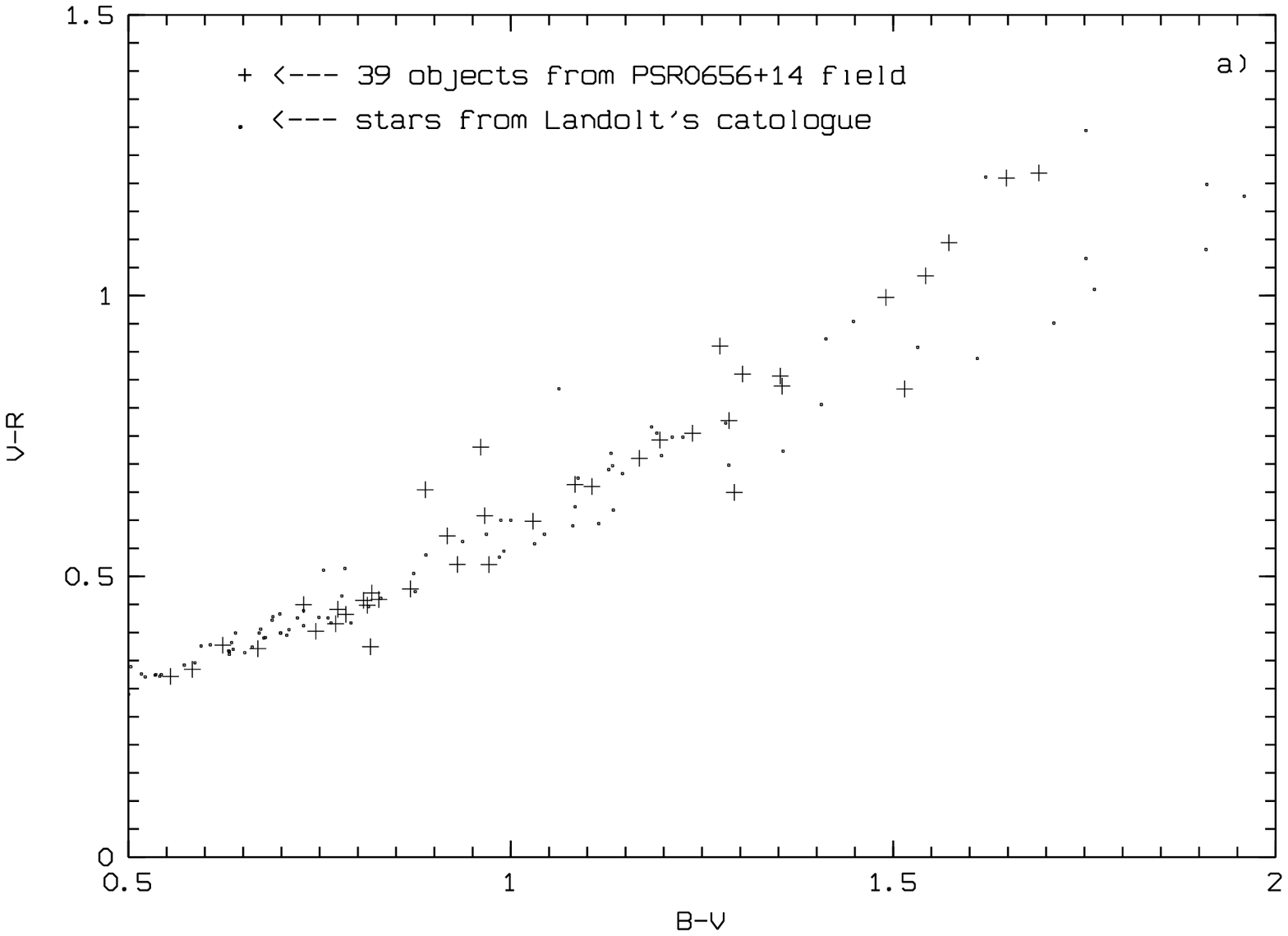,width=8.5cm,%}}\par
	bbllx=30pt,bblly=30pt,bburx=570pt,bbury=425pt,clip=}}\par
\vspace{-6.25cm}
\hspace{8cm}
    \vbox{\psfig{figure=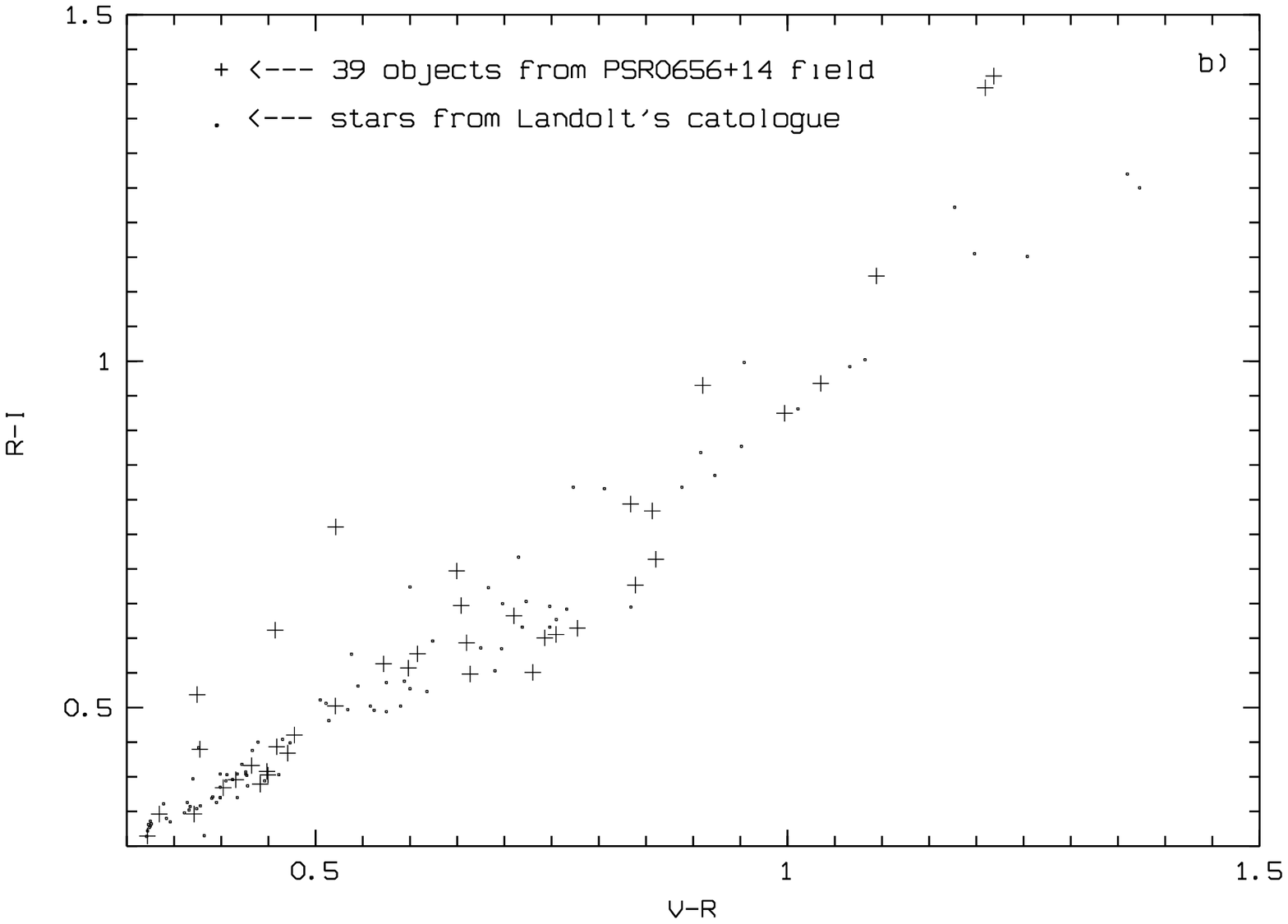,width=8.5cm,%}}\par
	bbllx=30pt,bblly=30pt,bburx=570pt,bbury=425pt,clip=}}\par
}
\caption{{\bf a)} $V-R$ vs. $B-V$ diagram for stars of the PSR~B0656+14 field.
Crosses denote the color indices for the field objects.
Points show the color indices for stars from the Landolt catalog.
{\bf b)} $R-I$ vs. $V-R$ diagram for the field stars.
}
\label{BVRI}
\end{figure*}

\begin{figure*}
\centerline{
\vbox{\psfig{figure=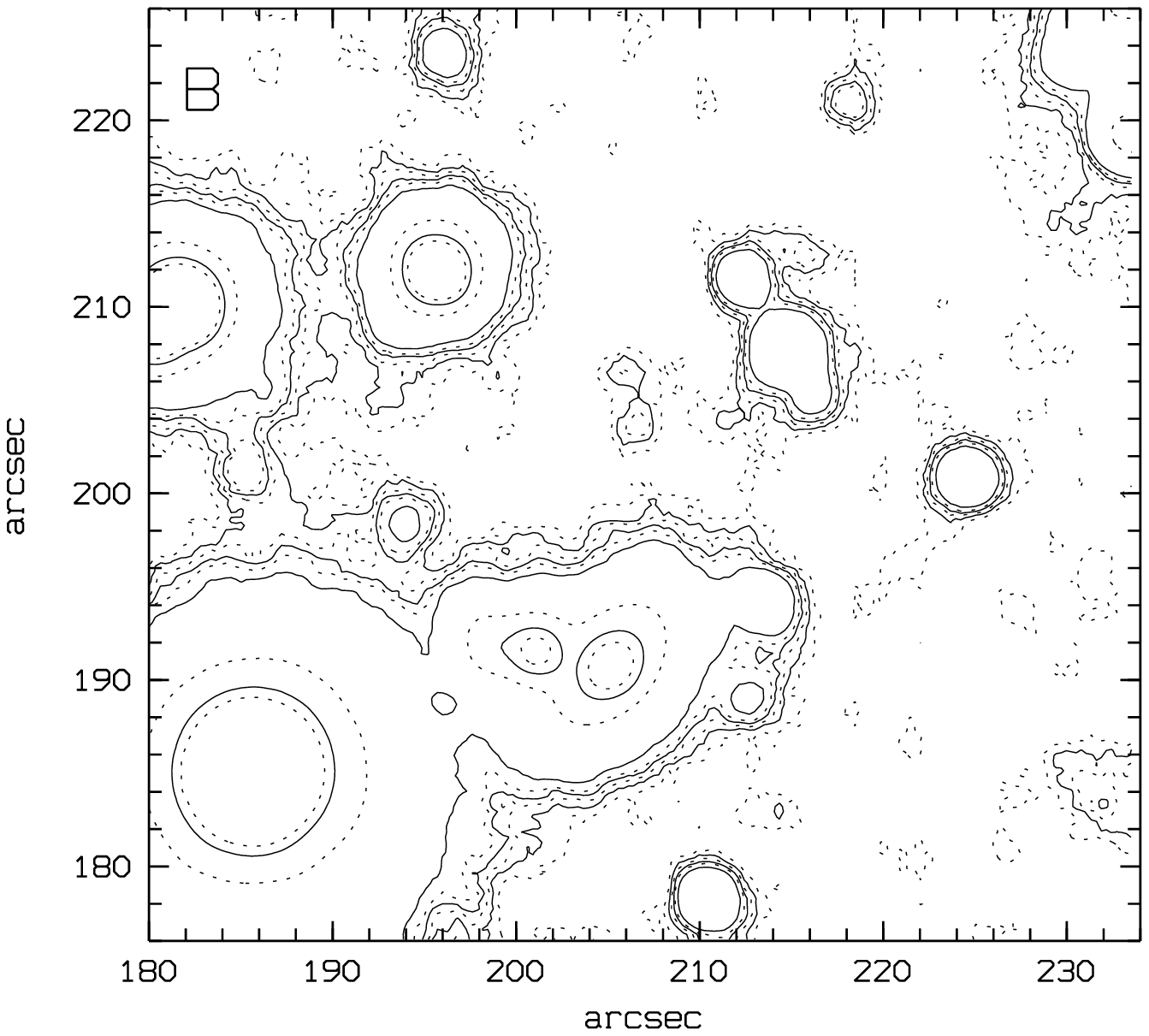,width=11.cm,%
 bbllx=35pt,bblly=50pt,bburx=445pt,bbury=435pt,clip=}}}  \par
\vspace{-1.cm}
\centerline{
\vbox{\psfig{figure=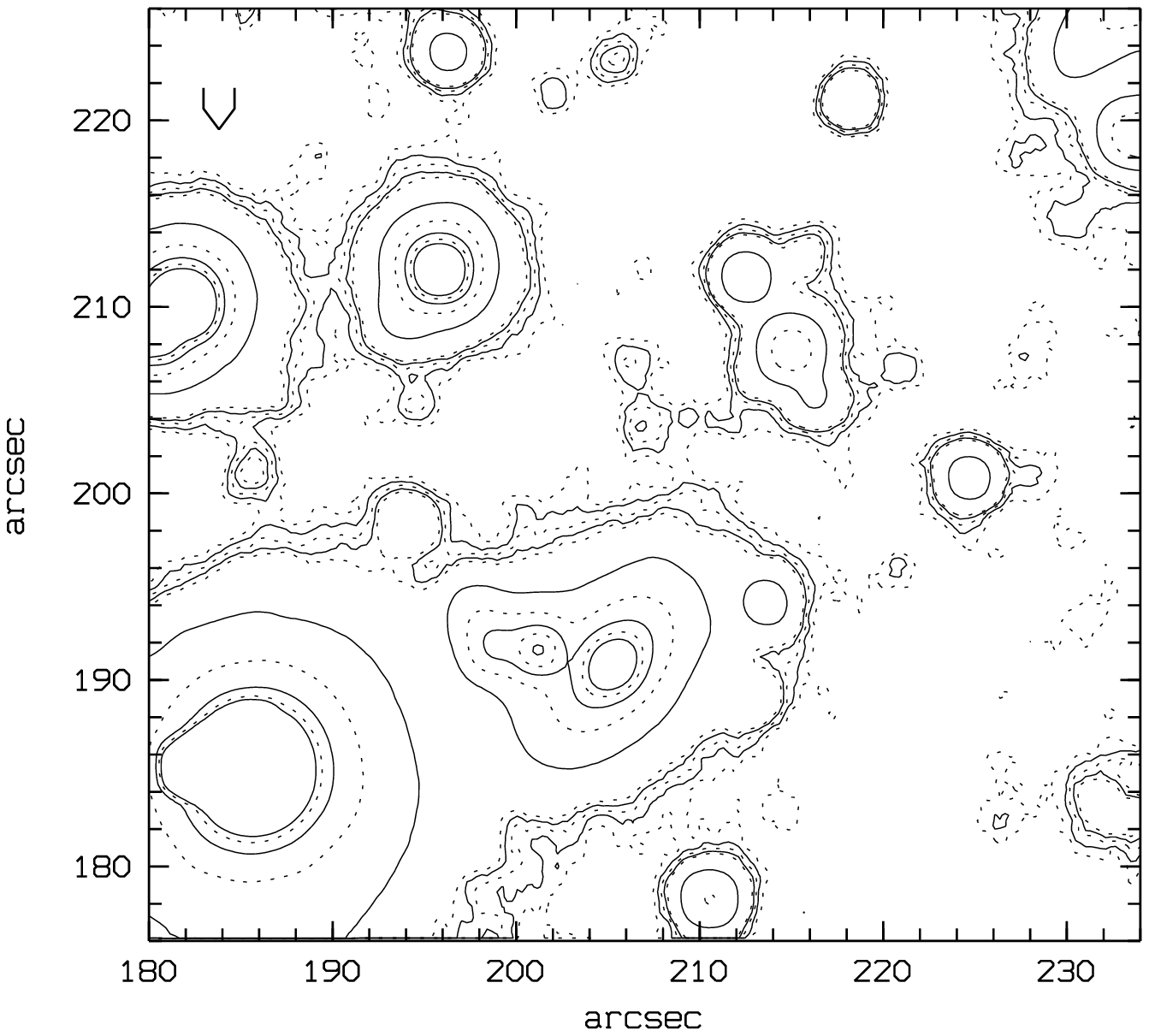,width=11.cm,%
 bbllx=35pt,bblly=50pt,bburx=445pt,bbury=435pt,clip=}}}  \par
\caption{Contour plots of the PSR~B0656+14 vicinity in the {\it B}
and $V$ filters.
The size of each fragment is $50^{\prime\prime}\times54^{\prime\prime}$.
}
\label{OTarBV}
\end{figure*}

\begin{figure*}
\centerline{
\vbox{\psfig{figure=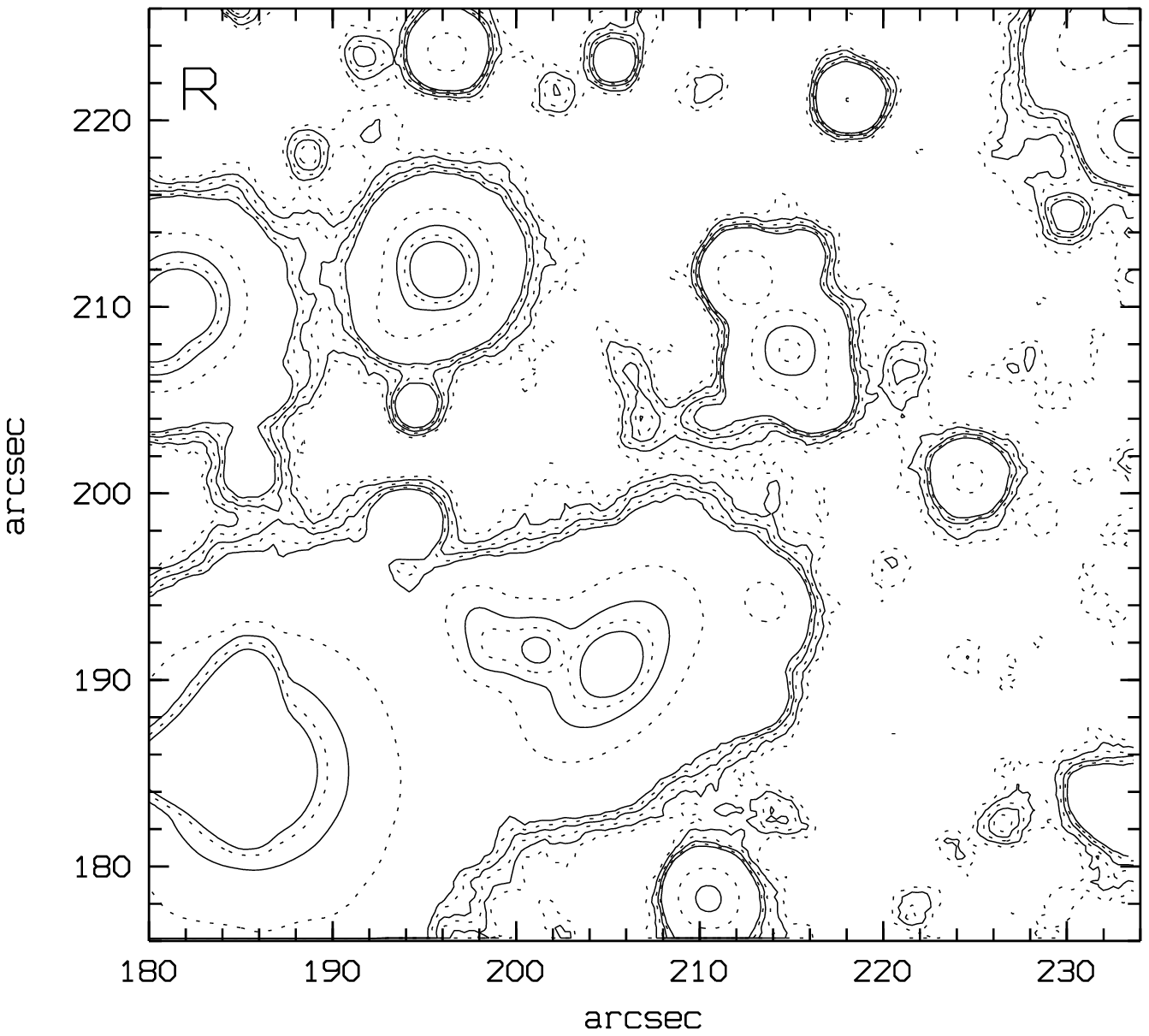,width=11.cm,%
 bbllx=35pt,bblly=50pt,bburx=445pt,bbury=435pt,clip=}}}  \par
\vspace{-1cm}
\centerline{
\vbox{\psfig{figure=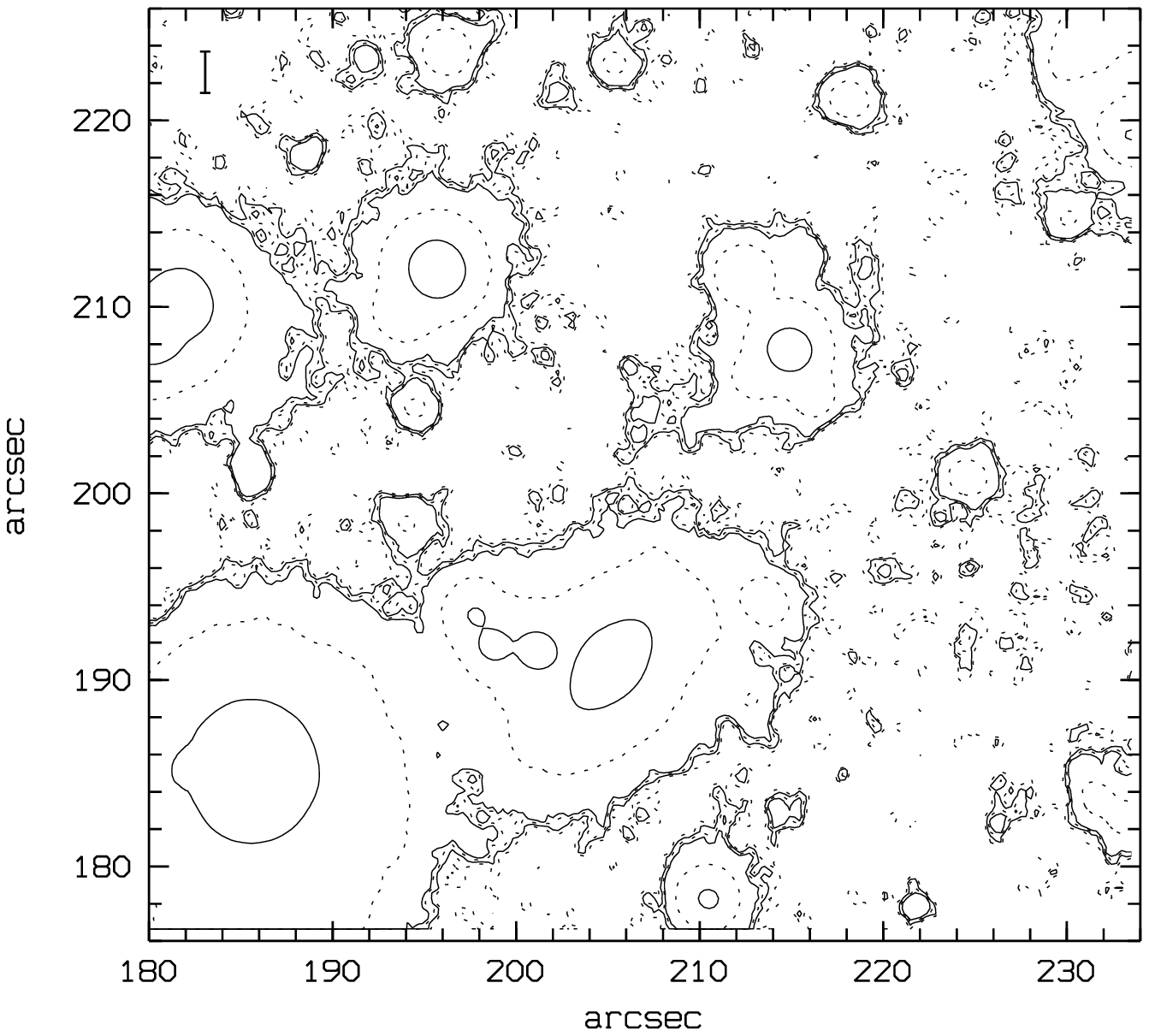,width=11.cm,%
 bbllx=35pt,bblly=50pt,bburx=445pt,bbury=435pt,clip=}}}  \par
\caption{Contour plots of the PSR~B0656+14 vicinity in {\it R}
and $I$ filters.
The size of each fragment is $50^{\prime\prime}\times54^{\prime\prime}$.
}
\label{OTarRI}
\end{figure*}

\subsection{Photometry}

 Photometric referencing was carried out by observing four standard
stars from the Landolt catalog (1992): 112822, GD71 and RU 149 (8 stars) and
star 2015 from Weis (1991).
Photometric conditions
remained stable during the whole night, with an
insignificant increase of absorption at the blue part of
the spectrum
at the end of the observations. For the extinction factors, we used average
values  of the observatory: $k_{b}=0.34$, $ k_{v}=0.21$, $ k_{r} =0.15$, $k_{i} =0.10$
(in stellar magnitudes).
From observations of the standard stars we derived the following
transition equations from the instrument system
(magnitudes {\it b, v, r, i}) to the Cousins system
(magnitudes {\it B, V, R, I}):

\begin{eqnarray}
 B-b& =& 26.14 + 0.16 (b-v) \nonumber \\
 V-v& =& 26.30 - 0.11  (b-v)  \\
 R-r& =& 26.62 - 0.02  (v-r) \nonumber \\
 I-i& =& 25.77 + 0.07  (r-i)~. \nonumber
\end{eqnarray}
The instrumental magnitudes were calculated as
\begin{equation}
b=-2.5\log(F_b/t_{\rm exp})-\Delta_a m_b - k_b/\cos Z~,
\end{equation}
etc, where $F$ is the DN value in a given aperture,
$t_{\rm exp}$ is the exposure time, $\Delta_a m$ is
the correction for the finite aperture
(obtained from the Point Spread Function [PSF] of bright stars),
and $k/\cos Z$ is the
correction for the atmospheric extinction.
Errors of the numerical coefficients in eq.~(1) do not exceed $\pm 0.01$.

The errors of the stellar magnitudes can be determined
by measuring the signal-to-noise ($S/N$) ratios:
\begin{equation}
	    \Delta m = - 2.5 \log(1\pm\frac{1}{S/N}),
\end{equation}
\begin{equation}
	  \frac{S}{N} = \frac{F\sqrt{Gain}}{\sqrt{F+Sky\times N_{\rm pix}}}~,
\end{equation}
where
$N_{\rm pix}$ is the number of pixels
in the aperture chosen,
{\it Sky} is the value of the local background in DN per pixel,
and $Gain=2.3e^-/$DN is the gain.
 The CCD readout noise is $\sim 8 e^-$ or $4-5$~DN and
one  was negligible in our observations.

Making use of eqs.~(1)--(4), we measured the magnitudes
of relatively bright stars
of the pulsar field (individual images were summed for each filter). The Cousins
magnitudes of 11 of these stars
(denoted by the numbers 1 through 11 in Fig.~1),
together with  the effective wavelengths and widths of the
Cousins filters, are presented in Table 3. Figure 2 shows the
comparison of
the color indices
for 39 objects of the pulsar field,
from $V=17.5$ to $V=22.5$,
with the color indices
for stars from the Landolt catalog.
The color-color dependences are close to each other,
which attests both low
absorption in the Galaxy for the direction towards
PSR~B0656+14 and the absence
of systematic shifts in our photometric system.

\subsection{Optical counterpart of PSR~B0656+14}

Contour plots of the pulsar vicinity in all the 4 filters
obtained from smoothed
($3\times3$ pixels)
images are shown in Figs. 3 and 4
%(На  scheme in Fig.6. также показаны )
The first  five
isophotes (for lower brightnesses) correspond to
the following values above the background:
$Sky\, +\, n \sqrt{Sky/N_{\rm pix}}$, with $n=1,\ 2,\ 3,\ 4,\ 5$,
where $N_{\rm pix}=28$ corresponds to the
aperture diameter of 6 pixels ($1\farcs6$),
and the value of $Sky$ is that for the background
near the pulsar.
The isophotes for higher brightnesses are scaled arbitrarily to
demonstrate
brighter details.
It is seen from Figs.~3 and 4 that 5 objects are detected
within $\lapr 5^{\prime\prime}$ around the expected pulsar position.
The position of one of them (``o1,PSR'' in Table 2 and Fig.~6)
coincides, within the budget of astrometry errors,
with the radio position,
so that this object is the most plausible candidate
for the optical counterpart of PSR~B0656+14. The nearest object ``o2''
(detected in the {\it I} filter only) lies to the North at a distance of
$\approx 2^{\prime\prime}$.
 Since the objects in the near vicinity are not seen simultaneously
in all 4 filters, we have shown them, for the sake of convenience,
in the separate scheme of Fig.6.

\begin{figure*} [t]
\centering{
\vbox{\psfig{figure=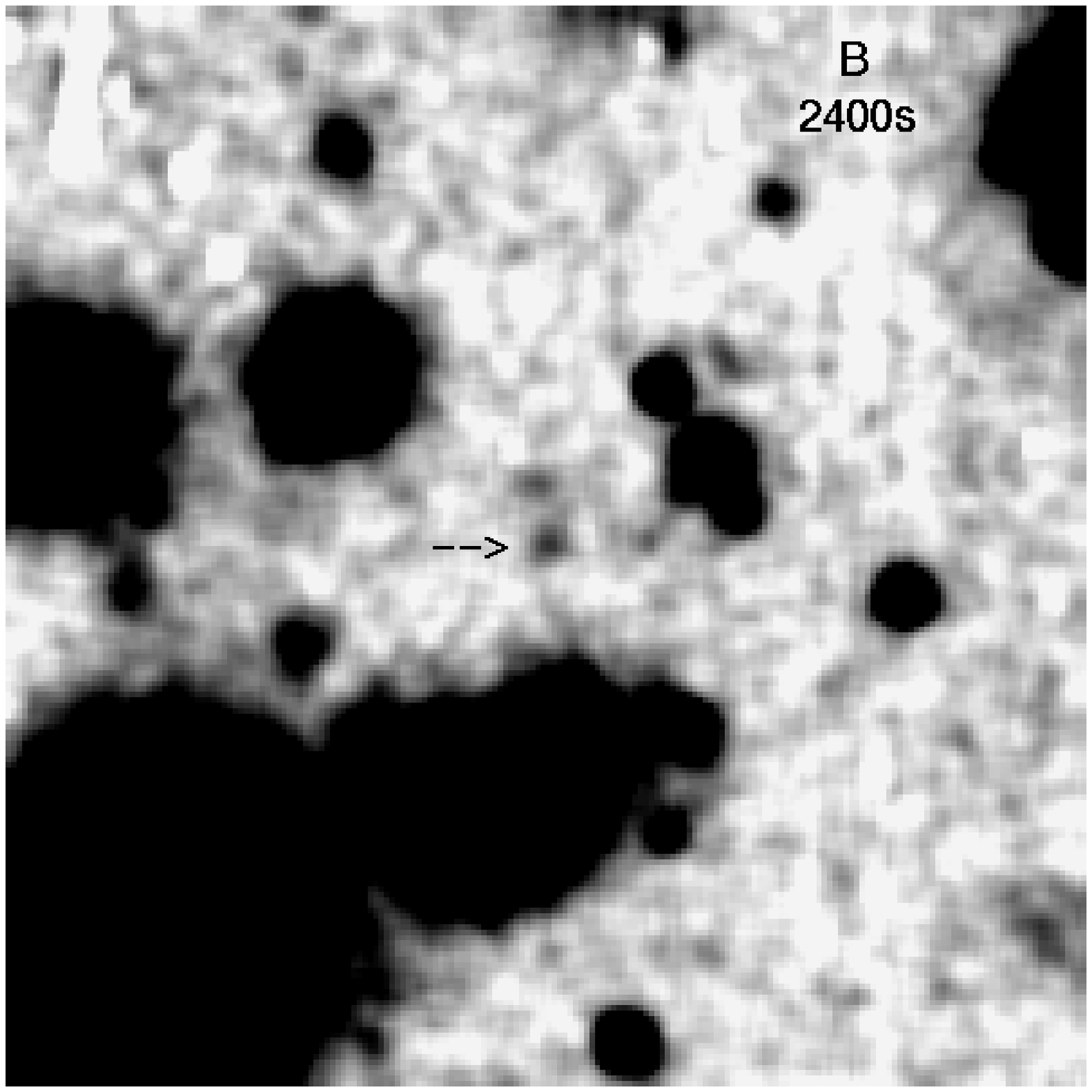,width=9cm,%
 bbllx=72pt,bblly=300pt,bburx=520pt,bbury=750pt,clip=}}  \par
  \vspace*{-9cm}\hspace*{9.015cm}
\vbox{\psfig{figure=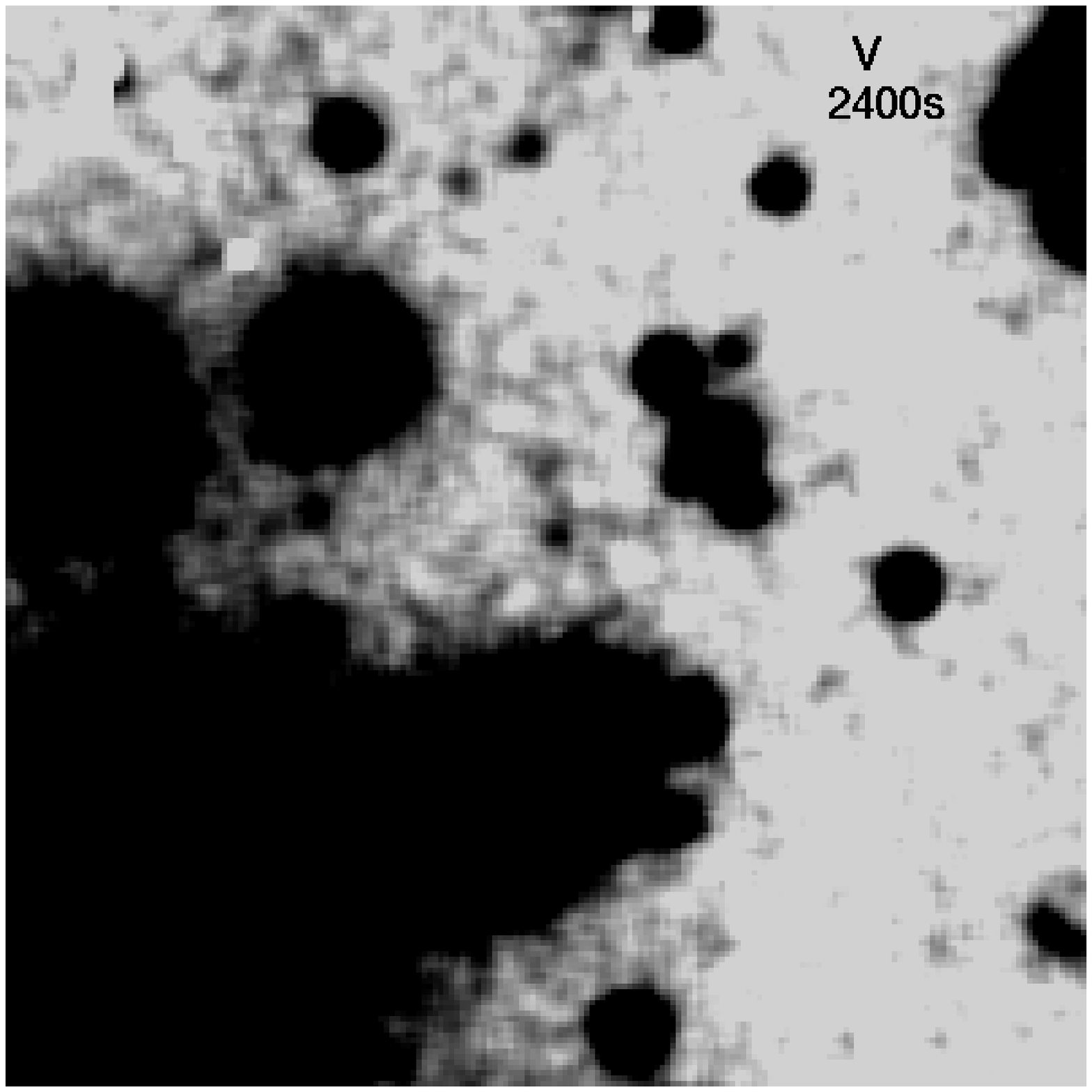,width=9cm,%
 bbllx=72pt,bblly=300pt,bburx=520pt,bbury=750pt,clip=}}  \par
\vbox{\psfig{figure=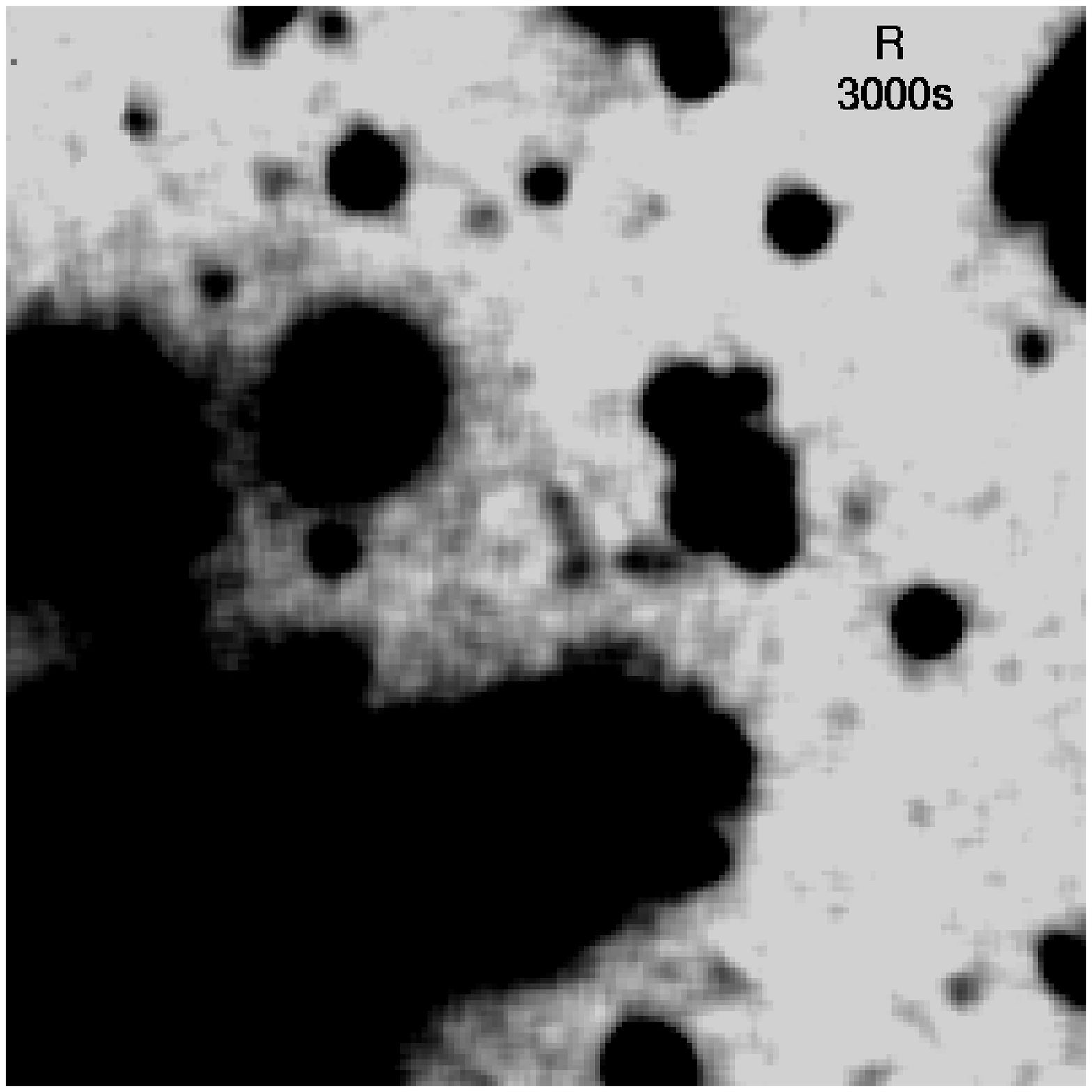,width=9cm,%
 bbllx=72pt,bblly=300pt,bburx=520pt,bbury=750pt,clip=}}  \par
\vspace*{-9cm}\hspace*{9.015cm}
\vbox{\psfig{figure=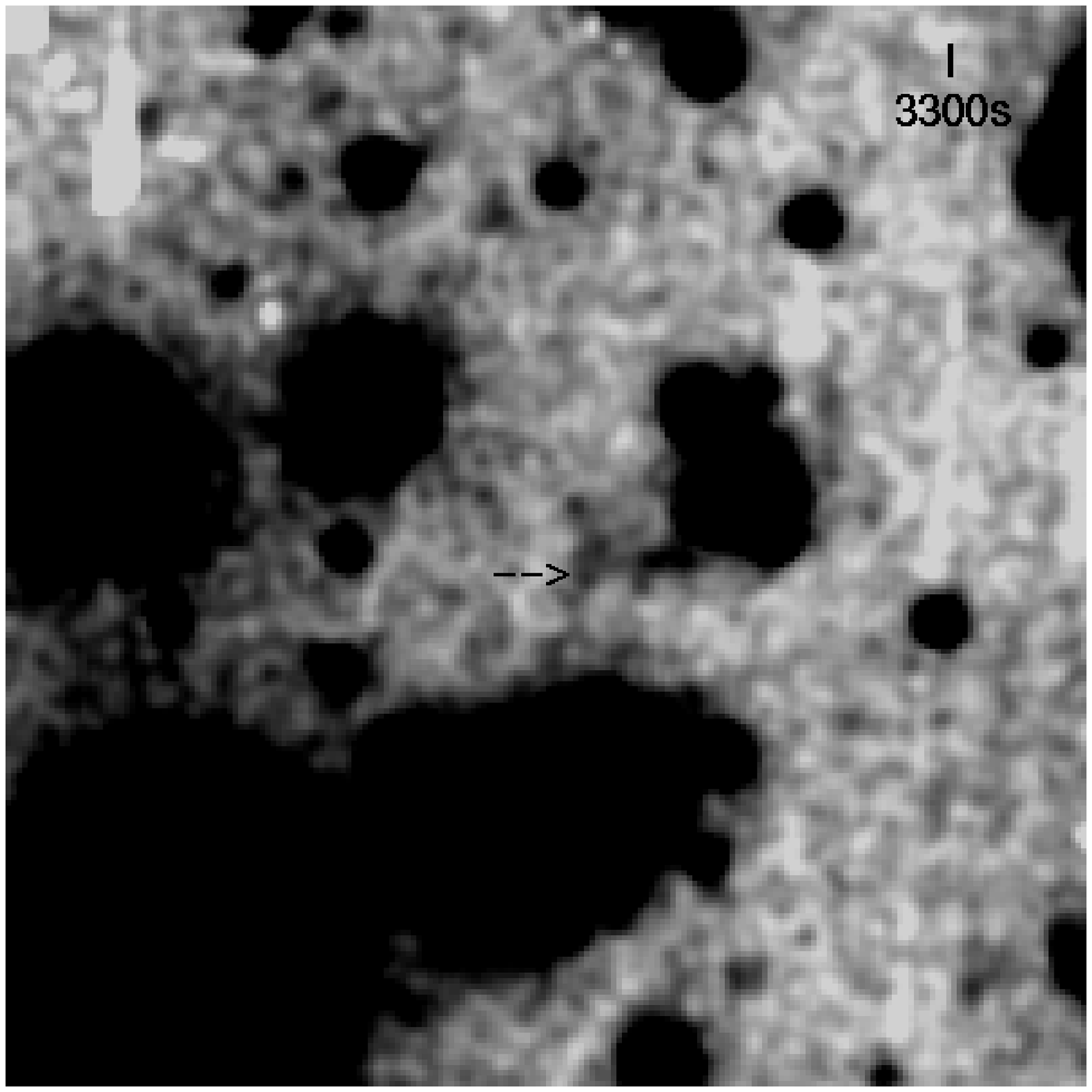,width=9cm,%
 bbllx=72pt,bblly=300pt,bburx=520pt,bbury=750pt,clip=}}  \par
}
\label{OTar1}
\caption{Vicinity of the pulsar PSR~B0656+14 in the {\it B, V, R, I} filters.
The size of each fragment is  $54^{\prime\prime}\times54^{\prime\prime}$.
The optical counterpart of the pulsar (``o1,PSR'', cf.~Fig.6) is indicated
by arrows in the {\it B} and {\it I} images.
The nearest object (``o2'' in Fig.~6) lies to the North at a distance of
$\approx 2^{\prime\prime}$.
}
\end{figure*}

\begin{table*}
\caption[ ]{Photometry of the PSR~B0656+14 optical counterpart.}
\begin{flushleft}
\begin{tabular}{c|c|c|c|c|c|c|c|c|c|c}
\hline
  filter & $Sky/t_{\rm exp}$ & $R_{a}$ & $N_{\rm pix}$ & $S/N$ & $-2.5\log (F/t_{\rm exp})$ &
$\Delta_a$m & $k/\cos Z$ & $m_{\rm ins}$ &  $ m_{\rm cous}$ & Flux \\
		 & (DN/pix/s) & (pix)  &       &           &      &       &       & &  & ($\mu$Jy)               \\ \hline
		 &       &   3    & 28   & 5.6  &  +0.13  & $1.01$     &      &       &       &                \\
    $B$          &2.465  &   4    & 50   & 6.1  & $-0.24$ & $0.63$     & $0.43$ & $-1.30$ &  $24.85(+0.19,-0.16)$ & $0.46\pm 0.08$  \\
		 &       &   5    & 79   & 6.1  & $-0.42$ & $0.41$     &      &       &       &                \\ \hline
		 &       &   3    & 28   & 6.6  & $-0.29$ & $0.84$     &      &       &       &                \\
    $V$          &6.429  &   4    & 50   & 6.7  & $-0.58$ & $0.51$     & $0.26$ & $-1.35$ &  $24.90(+0.16,-0.14)$ & $0.40\pm 0.06$ \\
		 &       &   5    & 79   & 7.1  & $-0.73$ & $0.33$     &      &       &       &                \\ \hline
		 &       &   3    & 28   & 7.9  & $-1.11$ & $0.81$     &      &       &       &                \\
    $R$          &18.207 &   4    & 50   & 9.8  & $-1.34$ & $0.48$     & $0.19$& $-2.09$ &  $24.52(+0.12,-0.11)$ & $0.47\pm 0.06$ \\
		 &       &   5    & 79   & 9.3  & $-1.60$ & $0.30$     &      &       &       &                \\ \hline
		 &       &   3    & 28   & 4.6  & $-1.02$ & $0.81$     &      &       &       &                 \\
    $I$          &29.922 &   4    & 50   & 4.2  & $-1.42$ & $0.49$     &  $0.12$ &$-1.95$  &  $23.81(+0.27,-0.21)$ & $0.71\pm 0.16$ \\
		 &       &   5    & 79   & 4.1  & $-1.67$ & $0.31$     &      &       &       &                  \\ \hline
\end{tabular}
\end{flushleft}
\label{Phot}
\end{table*}

The photometry of the pulsar candidate
(Table 4) was carried out for several aperture radii $R_a$,
to optimize the $S/N$ ratio.
The object ``o2'',
whose brightness in the $I$ filter is comparable to that
of the pulsar candidate,
starts to influence the photometry of the candidate
at $R_a \gapr 4$ pixels
($\gapr 1^{\prime\prime}$).
Most notations of Table 4 have been described above.
The columns $m_{\rm ins}$ and $m_{\rm cous}$ are for the magnitudes
in the instrumental and Cousins systems, respectively.
The last column shows the source flux (in $10^{-29}$ erg cm$^{-2}$
s$^{-1}$ Hz$^{-1}$) obtained from the Cousins magnitudes with the aid
of the data on $\alpha$Lyr by Fukigita et al.~(1995)

\subsection{Objects in near vicinity
%($\lea 5^{\prime\prime}$) of
of the PSR~B0656+14 counterpart}

The coordinates of the
5 objects
within $\lapr 5^{\prime\prime}$ of the expected pulsar
position
(see Figs. 5 and 6)
are presented in Table 2.

The object ``o1,PSR'', whose
brightness distribution is consistent with that of
a point source,
is the only
probable candidate for the
pulsar counterpart, based upon both its position and photometry.

\begin{figure}
\centerline{
    \vbox{\psfig{figure=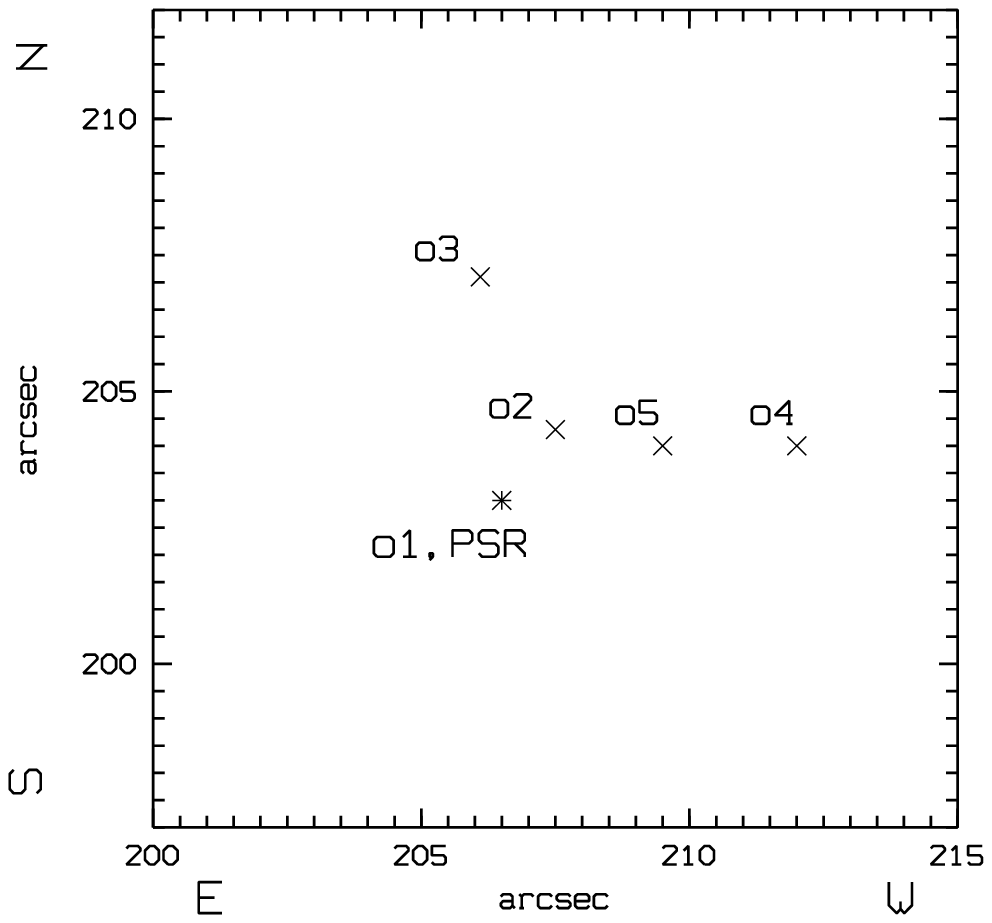,width=8.5cm,%}}\par
	bbllx=30pt,bblly=140pt,bburx=340pt,bbury=430pt,clip=}}
}
\caption{
 Objects from
nearest vicinity
of the pulsar are indicated whose
coordinates are given in Table 2, and
estimates of
stellar magnitudes are given in Table 5.
Coordinates
(in arcseconds) correspond to those in Figs.~3 and 4.}
\label{Shema}
\end{figure}

\begin{figure}
   \centering{
    \vbox{\psfig{figure=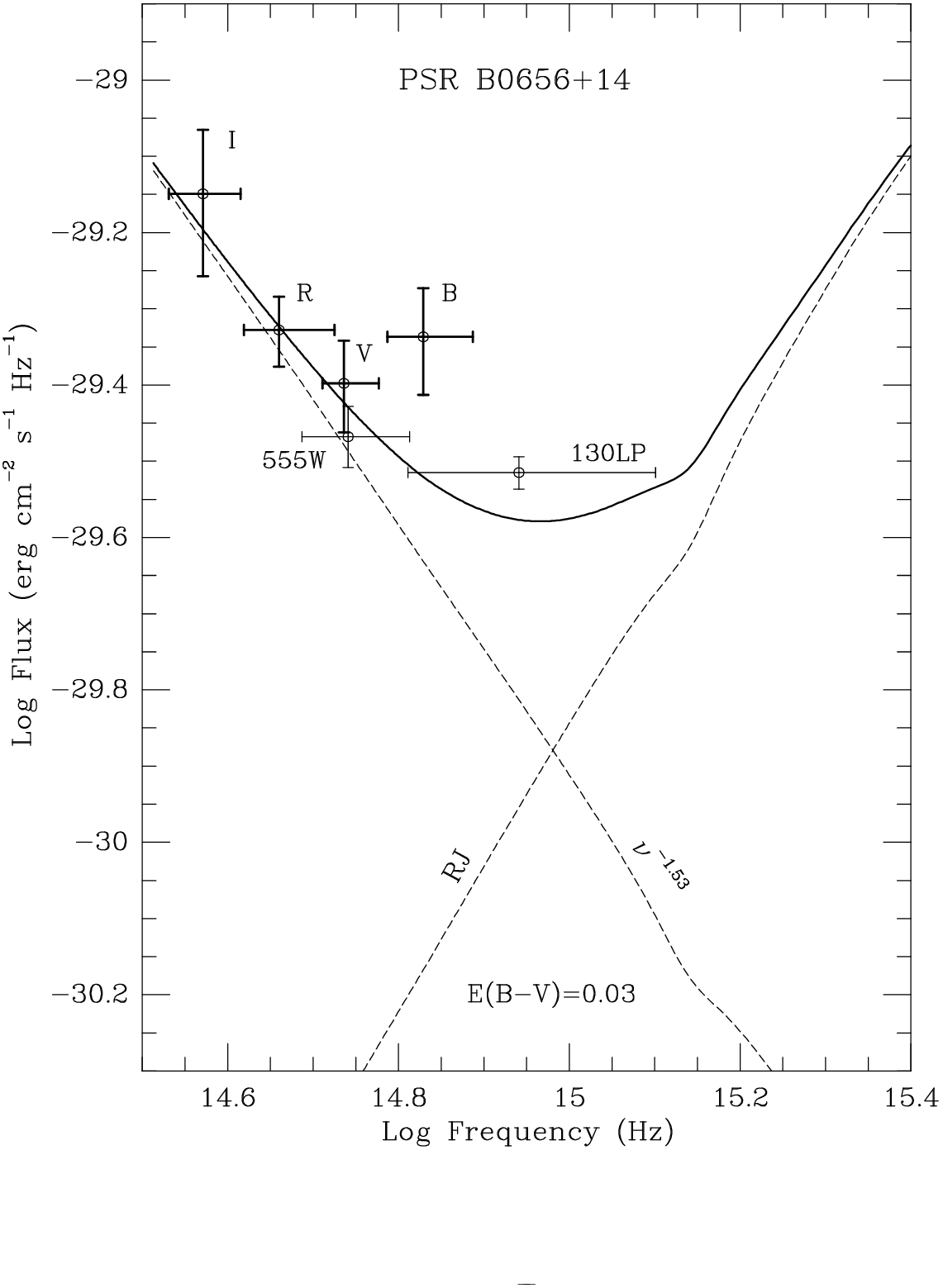,width=8.5cm,%}}\par
	bbllx=30pt,bblly=75pt,bburx=580pt,bbury=735pt,clip=}}\par
}
\caption{Fluxes observed from the PSR~B0656+14 optical counterpart
in different spectral bands. The crosses $B$, $V$, $R$ and $I$
show our results; the crosses 130LP and 555W
show the fluxes detected with the $HST$
FOC/F130LP by Pavlov et al.~(1996a) and WFPC2/F555W
by Mignani et al.~(1997).
Dashed crosses
show the fluxes obtained
by Caraveo et al. (1994; $V$ band)
and
Kurt et al. (1997a; $B$ band).
The 3$\sigma$ upper limits on the flux from undetected sources
in the {\it B,V,R,I} filters are shown
for the aperture diameter of 6 pixels ($\approx 1\farcs6$).
The solid line shows the best fit of the fluxes in the 6 bands
with the two-component model (power law + Rayleigh-Jeans spectrum;
eq.~[5]) for the color excess $E(B-V)=0.03$. The dashed lines
are the components of this model.
}
\label{spectr}
\end{figure}

The object ``o3''
is seen in all our {\it B,V,R,I} images,
as well as in the
contour plots of Fig.~2 of
Caraveo et al. (1994a).
It has been also detected as a faint extended object in
the $HST$  F130LP image
(Pavlov et al. 1996a).
Moreover, it can be seen even in our first {\it B} images
(Kurt et al. 1997a)
as a
background enhancement exactly at the same position.
Thus,
the presence of this faint (extended?)
object at $\simeq 4\asec$ to the North of the pulsar
is firmly established.
The object ``o4''
has also been detected with
the ESO-3.6m
and NTT telescopes in the {\it V} filter (Caraveo et al. 1994a),
although it is difficult to distinguish
between the
objects ``o4'' and ``o5'' because of their obviously extended nature.
These objects are outside of the $7\farcs4\times7\farcs4$ frame of
the $HST$ FOC image presented by Pavlov et al.~(1996a).

The object ``o2'', closest to the pulsar, is reliably detected only
in the {\it I}  filter.
In other filters only a non-significant enhancement of the background
is seen at this position.
We suppose that the object ``o2'' is not seen
in the $HST$ FOC image
because it is too red to be detected in the F130LP band.

\begin{table}
\caption[ ]{Stellar magnitudes of objects around PSR~B0656+14.}
\begin{center}
\begin{tabular}{|c|c|c|c|c|}
\hline
 object          &   $B$        &   $V$          & $R$                  & $I$
    \\  \hline
    o2           &   -          &  -           &  -                  &  23.83
   \\
		 &              &              &                     &   (+0.2
1;-0.19)  \\ \hline
    o3           &$25.0$        & $25.4$       &  24.63              &  23.93
	      \\
		 &$ \pm0.2$     & $\pm0.4$     &  (+0.15;-0.13)      &   (+0.2
4;-0.20)      \\ \hline
    o4           & $25.5$       & $25.5$       &  24.59              &  23.50
	      \\
		 & $\pm0.35$    & $\pm0.5$     &  (+0.21;-0.17)      &  (+0.12
;-0.10)       \\ \hline
    o5           &   -          &  -           &  24.35              &  23.54
	      \\
		 &             &             &   (+0.14;-0.12)     &  (+0.20;-
0.17)       \\ \hline
\end{tabular}
\end{center}
\label{Phot1}
\end{table}

Table 5 gives
estimates for magnitudes of these objects.
These estimates and the distances of the objects relative to ``o1,PSR''
(Table 2)
allow us to conclude that these objects
do not affect the measured values of the {\it B, V, R} magnitudes
of the pulsar counterpart.
A minor effect  about 3\% from whole flux of the nearest object ``o2'' on the $I$ magnitude
determined for the pulsar counterpart with the aperture radius
of 3 pixels (diameter of $1\farcs6$) can be expected,
but it should not exceed the errors of our measurement.

\section{Discussion.}

Figure 7 shows the
UV-optical energy fluxes
in different bands
for the PSR~B0656+14 counterpart obtained with
%different instruments and
%published by May 1997.
the $HST$ and 6-m telescope.
It demonstrates that our results
%(solid
(crosses $B$, $V$, $R$ and $I$),
being  compatible with
previously obtained data, considerably extend redward the observed frequency
domain, fill in the gap between the visual and UV parts
of the spectrum, and provide more accurate fluxes in the $V$ and
$B$ bands. The overall spectrum firmly confirms a nonthermal
origin of the optical radiation suggested by Pavlov
et al.~(1996a). On the other hand, the relatively high value
of the F130LP flux makes very plausible that the thermal radiation
from the NS surface contributes noticeably into the UV part of the
spectrum. Parameters of the thermal
and nonthermal fluxes can be constrained by fitting the
observed spectrum with a two-component model (cf.~Pavlov et al.~1996a),
a sum of the
power-law and thermal spectra (the latter has the form of a
Rayleigh-Jeans spectrum at these frequencies for plausible
NS temperatures):
\begin{equation}
f(\nu)=\left[f_0 \left(\frac{\nu}{\nu_0}\right)^{-\alpha} + g_0
\left(\frac{\nu}{\nu_0}\right)^2\right]\times 10^{-0.4 A(\nu)}~,
\end{equation}
where $\nu_0$ is an arbitrary reference frequency (we chose
$\nu_0=8.766\times 10^{14}$~Hz, which corresponds to $\lambda_0=
3420$~\AA), $A(\nu)$ is the interstellar extinction (see, e.~g.,
Savage \& Mathis 1979), $f_0$ and $g_0$ are the values of the
nonthermal and thermal fluxes at $\nu=\nu_0$. The latter can
be expressed, for the chosen value of $\nu_0$, as
\begin{equation}
g_0=
3.116\times 10^{-31} G ~\frac{{\rm erg}}{{\rm cm}^{2}~{\rm s}~{\rm Hz}}~,
\qquad G=T_6 \left(\frac{R_{10}}{d_{500}}\right)^2,
\end{equation}
where $T=10^6 T_6$~K is the NS brightness temperature,
$R_\infty = 10 R_{10}$~km is the NS radius (as seen by a distant
observer), and $d=500 d_{500}$~pc is the distance to the pulsar.
According to Pavlov et al.~(1996a), a plausible value for
the color excess, which determines the extinction curve $A(\nu)$,
is $E(B-V)=0.03$ ($A_V=0.09$). We fixed $E(B-V)$ at this level
and fitted the fluxes in the 6 bands ($B$, $V$, $R$, $I$, F555W
and F130LP) with eq.~(5), varying $f_0$, $\alpha$ and $G$.
The best-fit model (minimum $\chi^2=4.69$ for 3 degrees of freedom)
corresponds to $f_0=1.75\times 10^{-30}$~erg~cm$^{-2}$~s$^{-1}$~Hz$^{-1}$,
$\alpha=1.53$ and $G=4.1$. The best-fit spectrum
is shown in Fig.~7 with the solid line, and the best-fit thermal
and power-law components with dashed lines.

\begin{figure}[t]
   \centerline{
    \vbox{\psfig{figure=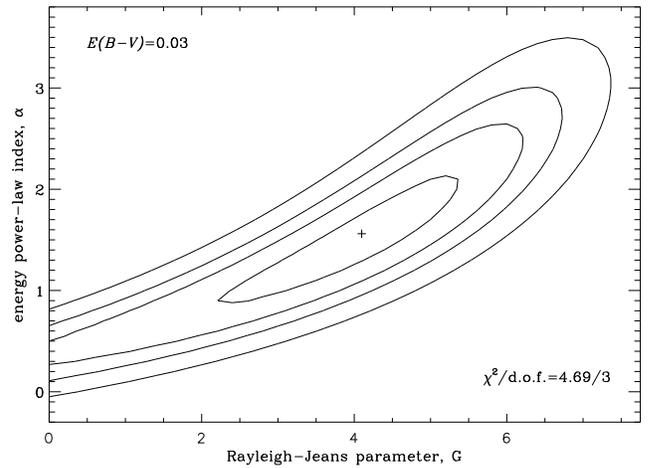,width=10.2cm,%}}\par
	bbllx=30pt,bblly=335pt,bburx=580pt,bbury=730pt,clip=}}
}
\label{fit}
\caption{Confidence levels (20, 68, 90 and 99\%) for fitting of the
observed spectrum of PSR~B0656+14 with the two-component model (eq.~[5])}
\end{figure}

The value of $G$ can be compared with what is expected from
extrapolations of the soft X-ray spectral flux observed with
the $ROSAT$ and $ASCA$ (Finley, \"Ogelman \& Kizilo\u{g}lu 1992;
Anderson et al.~1993;
Greiveldinger et al.~1997) to the optical range. The result of
this extrapolation depends on the model assumed to fit the
X-ray spectrum. The blackbody fit of the $ROSAT$ spectrum
yields $T_6=0.9\pm 0.04$ for $R_{10}/d_{500}=1.0\pm 0.2$ (Finley et al.~1992),
which corresponds to the Rayleigh-Jeans parameter
 $G=0.9\pm 0.3$. A fit of the same data with
a set of hydrogen NS atmosphere models (Pavlov et al.~1995) gives,
for $B=4.7\times
10^{12}$~G, the radius-to-distance ratio
$R_{10}/d_{500}=2.3\pm 0.4$ and the {\it effective} temperature
$T_\infty^{\rm eff}=(5.3\pm 0.5)\times 10^5$K (Anderson et al.~1993).
Since the model spectrum
differs from the blackbody spectrum, there is no a universal relation between
the effective and brightness temperatures; however, the model spectrum
in the optical range closely reminds, and can be fitted with,
the Rayleigh-Jeans spectrum with $T_6\simeq 0.9 T_\infty^{\rm eff}$,
thus giving $G\simeq 2.6\pm 0.7$, almost thrice greater than
$G$ inferred from the
blackbody model.

We see that the best-fit value of $G$ obtained from the available
optical-UV observations exceeds the predictions of both the
blackbody and hydrogen atmosphere interpretations of the soft X-ray flux.
However, as Fig.~8 shows, the fitting parameter $G$ is constrained
poorly by our data: $0<G<6.2$ at a $1\sigma$ level. The main
reason of that is the lack of data in the far-UV range
where the thermal component gives the main contribution.
Thus, we can only conclude that the results are
consistent with the presence of a thermal component,
but we cannot distinguish between different NS surface
models and can only estimate upper limits for the Rayleigh-Jeans
parameter, $G<6.2$, 6.8 and 7.3 at the confidence levels of
68, 90 and 99\%. We expect that better constraints will
be obtained from recent $HST$ FOC observations of this pulsar
in the UV range (Pavlov, Welty \& C\'ordova 1997).

Figure 7 clearly demonstrates that the pulsar radiation in the IR-optical
range is predominantly of a nonthermal origin, in agreement with
the conclusion of Pavlov et al.~(1996a). An independent confirmation
of the nonthermal nature comes from the high pulsed fraction
observed by Shearer et al.~(1997) in the $B$ band.
Although the slope of
the nonthermal spectrum is not well constrained because of relatively
large errors (especially in the $I$ band):
$0.3<\alpha <2.6$ at a $1~\sigma$ level,
it looks much steeper than
those of the (much younger) Crab pulsar, $\alpha=-0.11\pm 0.13$
(Percival et al.~1993; see also Ransom et al.~1994).
% and PSR~B0540--69,
%$\alpha \sim 0.24$ (Nasuti et al.~1997; Middleditch et al.~1987).
The nonthermal component
of the younger Vela pulsar has the slope $\alpha \simeq 0.3$ in the
$VBU$ range, but its $R$ flux is about twice lower than the extrapolation
of the power law (Nasuti et al.~1997). Even more drastic difference
is seen between the shapes of the optical spectra of PSR~B0656+14 and
the older pulsar Geminga
($\tau =3\times 10^5$~yr) which looks very much alike B0656+14 in
the soft X-ray range.
Geminga's flux has a minimum at the $B$ band,
 is peaked at the $V$ band, and falls off
sharply redward, which allowed Bignami et al.~(1996) to suggest
the presence of a proton cyclotron line centered at $\lambda\simeq 6000$~\AA.
On the contrary, the flux of PSR~B0656+14 {\it grows}
redward of $V$, thus demonstrating that the properties of the nonthermal
IR-optical-UV radiation are neither the same in different pulsars
nor they have a clear correlation with the pulsar age.

The apparent excess of the $B$ flux with respect to acceptable
continuum fits (Fig.~7), responsible for the relatively large
value of the minimum $\chi^2$,
may indicate the presense of a spectral feature
at these wavelengths, perhaps similar to that reported
for Geminga. More observations are needed to verify whether
the feature is real. But at least three broad emission lines above 3600 A 
were recently detected indeed in a prism spectrum of PSR B0540-69 
with the FOS/HST by Hill et al. (1997). The authors connect directly
the presence of these strong emissions with a nebula surrounding B0540-69. 

The nonthermal component has also been observed
 in high-energy radiation from
PSR~B0656+14.
Greiveldinger et al.~(1997) reported a power-law component in
the X-ray spectrum of PSR~B0656+14, with  $\alpha = 0.5\pm 1.1$,
which dominates at $E\gapr 2$~keV. Ramanamurthy et al.~(1996)
presented an evidence for $\gamma$-ray emission ($E>100$~MeV),
with $\alpha=
1.8\pm 0.3$. Being extrapolated to the X-ray range, our best-fit
power-law component ($\alpha=1.53$)  gives the X-ray flux about 3 orders
of magnitude lower than observed by Greiveldinger et al., i.~e.,
the nonthermal flux should grow somewhere between the UV and X-ray
ranges. To merge with the X-ray power-law component at $E\sim 2$~keV,
our optical flux should have a slope $\alpha \sim 0.3 - 0.4$,
hardly compatible with our fit (see Fig.~8), and in this case the
spectrum should steepen at higher energies in order to be
compatible with the $\gamma$-ray data. It is, however, hard to
believe that the spectrum would have
the same slope in the optical
and high-energy ranges --- in fact, none of the four other pulsars
with investigated nonthermal optical components shows such behavior.

To put more stringent constraints on the parameters of the thermal
and nonthermal components, more observations are needed. Observations
in the IR range would be most useful for elucidating the slope $\alpha$ of
the nonthermal component, whereas far-UV observations would be
crucial to measure the Rayleigh-Jeans parameter $G$ proportional to
the temperature of the NS surface. Since $G$ and $\alpha$ are correlated
with each other (see Fig.~8), measuring one of these parameters with a higher
accuracy would immediately constrain the other parameter.
It would be also useful to obtain more detailed spectrum
around the $B$ band
in order to understand whether the excess we observed is associated
with a spectral line. Such observations are quite feasible with
both the $HST$ and large ground-based telescopes.

{\bf Acknowledgements:}
 The work of G.~G.~P. was partially supported by NASA through
grant GO-06645.01-95A from the Space Telescope Science Institute,
which is operated by the Association of Universities for Research
in Astronomy, Inc., under NASA contract NAS 5-26555, and through
NASA grant NAG5-2807.

\end{document}